\documentstyle[11pt]{article}

\catcode`@=11
\@addtoreset{equation}{section}
\catcode`@=12

\begin{document}

                \begin{titlepage}
\setcounter{page}{1}

\date{}
\title{\small\centerline{May 1993 \hfill DOE-ER\,40757-012}
\small\rightline{CPP-93-12}
\small\rightline{LTP-038-UPR}
\bigskip\bigskip
{\Large\bf Induced charge of neutrinos in a medium}\bigskip}

\author{\normalsize \bf Jos\'e F. Nieves\\
\normalsize \em Physics Department, University of Puerto Rico,
              Rio Piedras, PR 00931
\medskip \\
\normalsize \bf Palash B. Pal\\
\normalsize \em Center for Particle Physics,
 University of Texas, Austin, TX 78712, USA}

\maketitle
\vfill
                \begin{abstract}
Neutrinos can polarize a medium due to their weak interaction. This
can manifest itself as an effective induced charge of the neutrino. We
show how it is related to the Debye screening length in a plasma,
first using the results of the 1-loop calculation of the
neutrino electromagnetic vertex and then in general to
all orders in $e$ and to leading order in the Fermi coupling. We also
discuss how the results are modified if the neutrinos have mass --
either Dirac or Majorana type.
                \end{abstract}
\vfill
\centerline{PACS numbers: 14.60.G}

\thispagestyle{empty}
                \end{titlepage}

\section{Introduction}
The properties of neutrinos that propagate through a medium
have been the subject of great interest
in the recent literature.  This has been motivated
by the attractive suggestion that  the solar neutrino
problem  \cite{palrev} can be solved by the resonant oscillation mechanism
\cite{MiSm86}, which hinges on the characteristics of the neutrino
propagation in a background medium \cite{Wol78}. Inspired by its potentially
important effects, the neutrino interactions in a
material environment have also been studied in some detail
 \cite{NiPa89,SeSm89,OrSe87,OPSS87,DNP89,DNP90,GKL91,Saw92,GKLL92}. Of
primary interest along these lines is the study of the
electromagnetic interactions of neutrinos in a medium.
A classic problem in this field is the decay of a plasmon into
$\nu\bar\nu$ pairs  \cite{ARW63}, which has received considerable
attention recently \cite{DNP89,Saw92,BrSe93}. It has also been pointed
out that the rate of decay of a massive neutrino into a lighter
neutrino and a photon increases tremendously \cite{DNP90,GKL91} in
matter, as a consequence of the fact that the GIM mechanism is not
operative in a medium with electrons but no muons or taons.

The study of the electromagnetic interactions of neutrinos in a
medium, as well as the results and conclusions mentioned above, are
based on the 1-loop calculation of the effective electromagnetic
vertex of the neutrino, which has been performed to the leading order
in the Fermi constant  \cite{DNP89} using the methods of ``Quantum
Statistical Field Theory''\footnote{More often, this is called the
``Finite Temperature Field Theory'', but the name is misleading
because the methods are also applicable at zero temperature with a
finite density of particles in the background.}. The implications of
this calculation have been only partially explored. In this article,
we show that the results of Ref. \cite{DNP89} imply that the neutrino
acquires a small effective charge in a medium. This point was realized
by Oraevsky and Semikoz  \cite{OrSe87} using methods of plasma physics
before the calculation of the $\nu\nu\gamma$ vertex was performed. On
the other hand, our method is entirely different, hopefully easier to
follow for a particle physicist, and it brings out some interesting
points which are not easy to see in the method used by Oraevsky and
Semikoz \cite{OrSe87}.  In addition, we also calculate the induced
charge for neutrinos which are massive, distinguishing between the
cases of Dirac and Majorana masses.

The plan of the paper is as follows. In Section 2 we
establish notation and the definition of the induced charge that
is used in the rest of the paper, and show why the
neutrino can acquire an induced charge in the medium but not in the
vacuum.  Then, using field-theoretic methods, we will show that the
neutrino electromagnetic vertex is related to the photon self-energy
in the medium and, in particular, the neutrino induced charge is
related to the Debye screening length in a plasma.
This relation is derived
to 1-loop order in Section 3, and then in Section 4 we show that it is
valid to all orders in $e$ (and to first order in the
Fermi coupling).  In Section 5 we use this relation
to find an expression for the induced charge of the neutrino,
and in Section 6 we show that the magnitude of the induced charge
depends on whether the neutrinos are massless or massive and,
in the latter case, on whether the mass is Dirac or Majorana
type. Finally, using the results of Ref. \cite{DNP89}, we will
estimate the induced charge in some particular backgrounds.

\section{Electromagnetic vertex of the neutrino and the definition of
the induced charge}
                       We begin by understanding the reason why the
neutrino can acquire an induced charge in the medium but not in the
vacuum.  The off-shell electromagnetic vertex function
$\Gamma_\lambda$ is defined in such a way that, for  on-shell
neutrinos, the $\nu\nu\gamma$ amplitude is given by
                \begin{eqnarray}
M = -i\overline u(k') \Gamma_\lambda u(k) A^\lambda(q)\,,
\label{defGamma}
                \end{eqnarray}
where
                \begin{eqnarray}
q \equiv k - k'
                \end{eqnarray}
is the momentum carried by the photon. In general, $\Gamma_\lambda$
depends on $k$ and $k'$ or, equivalently, on $k$ and $q$. For
neutrinos in a medium $\Gamma_\lambda$ depends also on the parameters
characterizing the medium. For homogeneous and isotropic media, to
which we will restrict ourselves, there is only one such parameter,
viz. the velocity 4-vector of the background medium  $v^\mu$.

There are two important consequences of the fact that the external
(neutrino) lines in the Feynman diagram for the $\nu\nu\gamma$
amplitude are neutral.  Firstly, $\Gamma_\lambda$ satisfies
                \begin{eqnarray}
\label{currconservation}
q_\lambda \Gamma^\lambda = 0
                \end{eqnarray}
for all values of $q$.  It is important to realize that
for neutrinos Eq. (\ref{currconservation}) holds for arbitrary values
of $q$, and not just when $k$ and $k'$ are on shell.
If the fermion lines in the diagram were to correspond to a charged
particle (e.g., the electron), then the analogous relation
in that case involves terms in the right-hand side involving
the inverse propagators corresponding to the external fermions.
In that case Eq. (\ref{currconservation}) does not hold
for arbitratry values of $q$, but only when both $k$ and $k'$
are on shell.

The other important consequence
of the fact that the external lines are neutral is that $\Gamma_\lambda$
is well defined in the limit $q^\mu \to 0$.  The reason is
that the photon vertex must be connected to a pair of internal
lines of the diagram. If one of these lines is assigned the loop
momentum $p$ which is integrated over, the other line will carry a
momentum $p \pm q$. The
propagator of this second line will involve the factor
                \begin{eqnarray}
{1 \over (p \pm q)^2 - m^2} =
{1 \over q^2 \pm 2 p\cdot q + (p^2 - m^2)} \,,
                \end{eqnarray}
where  $m$ is the mass of the internal line. However, since $p^2 \neq
m^2$ for any internal line, no singularity is produced for $q^\mu \to
0$.

{}From the two properties of $\Gamma_\lambda$ just discussed, we obtain
                \begin{eqnarray}
\Gamma_\lambda(q^\mu = 0) = 0 \,,
                \end{eqnarray}
which implies that the particle associated
with the external line does not acquire a charge in any order of
perturbation theory.
To see this explicitly, we will consider the matrix element of the
charge operator between two neutrino states with momenta:
                \begin{eqnarray}
k^\lambda  =  (E,\vec k), \qquad
k^{\prime\lambda} = (E,\vec k') \,.
                \end{eqnarray}
We use states with the same energy, because then
                \begin{eqnarray}
q \equiv k - k' = (0,\vec q)
                \end{eqnarray}
with $\vec q = \vec k - \vec k'$,
which corresponds to the static limit.  Denoting by $\rho(x)$ the
charge density operator, the effective charge is defined by the
equation
                \begin{eqnarray}
e_{{\rm eff}} \langle k'|k\rangle
& = & \int d^3x \langle k' \left| \rho(x) \right| k \rangle\nonumber\\
& = & (2\pi)^3 \delta^{(3)}(\vec q) \langle
k' \left| \rho(0) \right| k \rangle
\label{effch1}
                \end{eqnarray}
On the other hand,
                \begin{eqnarray}
\langle  k' \left| \rho(0) \right| k \rangle
= \overline u (k') \Gamma_0 (0,\vec q) u(k)\,,
\label{rhoGamma}
                \end{eqnarray}
where the notation $\Gamma_\lambda(q^0,\vec q)$ has been used to
indicate explicitly that we are considering the dependence of $\Gamma$
separately on the frequency and wavelength of the photon. Thus we
obtain
                \begin{eqnarray}
\label{basic}
e_{{\rm eff}} =
\frac{1}{2E} \overline u(k) \Gamma_0 (0,\vec q\to 0) u(k) \,,
                \end{eqnarray}
which is the basic equation to interpret our results, but it can be
cast in an elegant way. Introducing the spinor projection
matrix
                \begin{eqnarray}
S(k) \equiv u(k) \otimes \overline u(k) = {1 \over 2} (1 + \lambda
\gamma_5) \rlap/ k  \,,
\label{S(k)}
                \end{eqnarray}
where the last step is valid for massless Weyl spinors with
$\lambda=\pm 1$ being the helicity, we can rewrite Eq.\ (\ref{basic})
as
                \begin{eqnarray}
e_{{\rm eff}} &=&
{1 \over 2E} \; {\rm tr} \, [\Gamma_0 (0,\vec q\to 0) S(k)] \,,
\label{matrixeffch}\\
& = & {1 \over 4E} \; {\rm tr} \, \left[ \Gamma_0 (0,\vec q\to 0)
(1 + \lambda \gamma_5) \rlap/ k \right] \,,
\label{def:effch}
                \end{eqnarray}
where again the second step is valid for massless Weyl spinors.

Since $\Gamma_\lambda$ has a well defined limit as $q^0 \to
0$, we can make a Taylor expansion around $q^0 = 0$:
                \begin{eqnarray}
\Gamma_0 & = & G_0 + q^0 G_1 + O(q^{0^2})\,\nonumber\\
\vec\Gamma & = & \vec H_0 + q^0\vec H_1 +  O(q^{0^2})\,,
                \end{eqnarray}
where all the coefficients are independent of $q^0$.
Then from Eq. (\ref{currconservation})
it is easy to deduce that, in the limit $q^0\to 0$,
                \begin{eqnarray}
\label{q0limit}
\Gamma_0 & = & \vec q\cdot\vec H_1 + O(q^0)\,\nonumber\\
\vec\Gamma & = & q^0\vec H_1 +  O(q^{0^2})\,.
                \end{eqnarray}
Since $\vec H_1$ has a well defined limit
as $\vec q\to 0$, it follows that $\Gamma_0 = 0$ in this limit, and
from Eq. (\ref{def:effch}) we see that $e^{(\nu)}$ is zero in the
vacuum.

In the medium, Eq. (\ref{currconservation}) continues to hold for any
value of $q$, and the relations in Eq. (\ref{q0limit}) are also
valid, but $\vec H_1$ no longer has a well defined limit
as $\vec q\to 0$. There are several ways to understand
why this is so.  One of them is to notice that in the case of
the medium, some of the internal lines to which the photon
is attached are on shell because they correspond
to particles that are in the background.  Thus, the singularities
that are avoided in the case of the vacuum because the photon
is attached only to internal off-shell lines, reappear here. Therefore,
nothing prevents $\vec H_1$ to develop a singularity as
$\vec q\to 0$ of the form
                \begin{eqnarray}
\vec H_1 = \mbox{(constant)} \cdot \frac{\vec q}{{\vec q}\,^2}
                \end{eqnarray}
In such a case,  the constant appearing in this equation is
the value of $\Gamma_0$ in the limit $q^0=0,\,\vec q \to0$. Thus,
from the definition of the effective charge
in Eq. (\ref{def:effch}), it follows that  $e^{(\nu)}_{{\rm eff}}$ is
non-vanishing.

In what follows, we will show that this is exactly what happens, first
explicitly by using the 1-loop calculation of the neutrino
electromagnetic vertex and then by a general field-theoretic argument,
which extends the 1-loop result to all orders in $e$. This will be
done by showing that $\Gamma_\lambda$ is related to the photon
self-energy $\pi_{\lambda\rho}(q)$. Further, in the limit that we are
considering, $\pi_{\lambda\rho}(q^0 = 0,\vec q\to 0)$ is related to
the Debye screening length, which allows us to establish the relation
between the latter quantity and the neutrino induced charge.

\section{One-loop result}
                       We are interested in the regime where the
neutrino momenta are small compared to the masses of the $W$
and $Z$ bosons. Therefore, we can neglect the momentum dependence in
the $W$ and $Z$ propagators, which is justified if we are performing a
calculation to the leading order in the Fermi constant $G_F$. In this
approximation, the diagrams contributing to the electromagnetic vertex
then appear at the 1-loop level, and are shown in Fig.~\ref{f:1loop}.
Since the momentum dependence of the weak gauge bosons are neglected,
these two diagrams can be represented by the single diagram of
Fig.~\ref{f:4fermi}
with a four-fermion vertex. Let us denote the 4-fermion interaction
by
                        \begin{eqnarray}
                \label{Lweak}
{\cal L} {\rm _{int}^{(weak)}} =
-\sqrt 2 G_F \; [\bar \nu \gamma^\rho L\nu ] \;
[\bar f \gamma_\rho ({\cal A + B} \gamma_5) f ]
                        \end{eqnarray}
where $L={1\over2}(1-\gamma_5)$ is the projection operator for left
chirality, and $f$ stands for the electron field. We can then write the
amplitude of Fig.~\ref{f:4fermi} as
                        \begin{eqnarray}
-i \Gamma_\lambda &=& (-ie)(-iG_F\surd 2)(-1) \gamma^\rho L \;
\times \nonumber\\
&&\quad \int {d^4p \over (2\pi)^4} \;\mbox{tr}\,
\left[\gamma_\lambda \, iS_F(p) \gamma_\rho ({\cal A + B} \gamma_5) \,
iS_F(p - q) \,  \right] \,,
\label{Gamma}
                        \end{eqnarray}
where $iS_F(p)$ denotes the  propagator of the background particles
with momentum $p$, and $e$ is the charge of the electron. In
complicated systems, this propagator may be
complicated, and the integration over the momentum $p$ may have
unusual measure as well, but we do not need these explicitly for what
follows.

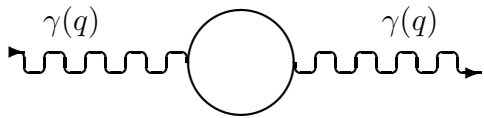
\begin{figure}
                \begin{center}
                \begin{picture}(70,25)(-10,20)
\thicklines
\multiput(15.5,30)(-6,0){4}{\oval(3,3)[t]}
\multiput(12.5,30)(-6,0){4}{\oval(3,3)[b]}
\put(-8.5,30){\oval(3,3)[tr]}
\put(-8,31.5){\vector(1,0){1}}
\put(0,36){\makebox(0,0){{\large$\gamma(q)$}}}
\put(25,30){\circle{16}}
\multiput(34.5,30)(6,0){4}{\oval(3,3)[b]}
\multiput(37.5,30)(6,0){4}{\oval(3,3)[t]}
\put(58.5,30){\oval(3,3)[bl]}
\put(59.5,28.5){\vector(1,0){1}}
\put(50,36){\makebox(0,0){{\large$\gamma(q)$}}}
                        \end{picture}
                \end{center}
\caption[]{\sf 1-loop diagram for the vacuum polarization of the
photon.}\label{f:vacpol}
                        \end{figure}

Now the interesting observation is that the contribution
from ${\cal A}$ in Eq. (\ref{Gamma}) is intimately related to
the vacuum polarization
of the photon which, at 1-loop, arises from the diagram in
Fig.~\ref{f:vacpol} and is given by
                        \begin{eqnarray}
i \pi_{\lambda\rho} (q) = (-ie)^2(-1)
\int {d^4p \over (2\pi)^4} \;\mbox{tr}\,
\left[ \gamma_\lambda \, iS_F(p) \,
\gamma_\rho \, iS_F(p - q) \right] \,.
\label{pi}
                        \end{eqnarray}
Therefore, Eq. (\ref{Gamma}) can be written in the form
                        \begin{eqnarray}
\Gamma_\lambda = -\frac{G_F}{\sqrt 2 e}\gamma^\rho(1 - \gamma^5)
({\cal A} \pi_{\lambda\rho} + {\cal B} \pi^5_{\lambda\rho})\,.
\label{Gampi}
                        \end{eqnarray}
where we have defined
                \begin{eqnarray}
i\pi^5_{\lambda\rho} = (-ie)^2(-1)\int {d^4p \over (2\pi)^4} \;\mbox{tr}\,
\left[\gamma_\lambda \, iS_F(p)
\gamma_\rho \gamma_5 \, iS_F(p-q) \,
\right] \,.
                \end{eqnarray}

The term proportional to $\pi^5_{\lambda\rho}$
does not
contribute to $\Gamma_0$. The reason is that
the trace contains a factor of $\gamma_5$, and therefore can
be non-zero only if there are at least four other $\gamma$-matrices
present. Since the terms in a fermion propagator have at most one
$\gamma$-matrix, the trace involves $\mbox{tr}\,(\gamma_5\gamma_\lambda
\gamma_\rho \gamma_\alpha \gamma_\beta) =
4i\epsilon_{\lambda\rho\alpha\beta}$. After the $p$ integration
this can yield only a term proportional to
$\epsilon_{\lambda\rho\alpha\beta}q^\alpha v^\beta$,
which does not contribute to the zeroth component of
$\Gamma_\lambda$.

\section{Generalization to higher orders}
                         The above result, based on the 1-loop
calculation of the photon self-energy and the neutrino vertex function,
can be generalized as follows.  The photon self-energy
is defined in general by
                \begin{eqnarray}
i\pi_{\lambda\rho} &= & (-ie)^2\int{d^4x e^{iq\cdot x}\left<  T\,
U^{{\rm (em)}}j_\lambda(x)j_\rho(0)\right> }\,,\nonumber\\
& = & (-ie)^2\int{d^4x e^{-iq\cdot x}\left<  T\,
U^{{\rm (em)}}j_\lambda(0)j_\rho(x)\right> }\,,
                \end{eqnarray}
where $j_\lambda$ is the electron current density
                \begin{eqnarray}
j_\lambda = \overline f\gamma_\lambda f\,,
                \end{eqnarray}
and
                \begin{eqnarray}
U^{{\rm (em)}} = \exp \left( i\int d^4z \,
{\cal L}{\rm _{int}^{(em)}} \right)
                \end{eqnarray}
with
                \begin{eqnarray}
{\cal L} {\rm _{int}^{(em)}} = -e \, j_\lambda A^\lambda\,.
                \end{eqnarray}
On the other hand, the neutrino vertex function of Eq.
(\ref{defGamma}) is defined by
                \begin{eqnarray}
\label{genvertex}
\Gamma_\lambda(k,k') = e \int{d^4x\,d^4y\, e^{-ik\cdot x}
e^{ik'\cdot y}\left<  T\, \exp \left( i\int d^4z \,
{\cal L} {\rm _{int}^{(total)}} \right)
\nu(y)j_\lambda(0)\overline\nu(x)\right> _a}\,,
                \end{eqnarray}
where
                \begin{eqnarray}
{\cal L} {\rm _{int}^{(total)}} = {\cal L} {\rm _{int}^{(weak)}} +
{\cal L} {\rm _{int}^{(em)}} \,.
                \end{eqnarray}
The subscript $a$ in Eq. (\ref{genvertex}) is used to indicate that
$\Gamma_\lambda$ is obtained from the above formula by amputating
the propagators corresponding to the external neutrino lines.
It is convenient to rewrite
${\cal L}_{{\rm int}}^{{\rm (weak)}}$, given in Eq. (\ref{Lweak}),
in the form
                \begin{eqnarray}
\label{Lweak2}
{\cal L}_{{\rm int}}^{{\rm (weak)}} = -\sqrt2 G_F \overline\nu_L
\gamma_\lambda \nu_L
\left[{\cal A}j_\lambda + {\cal B}j^5_\lambda\right]\,,
                \end{eqnarray}
where
                \begin{eqnarray}
j^5_\lambda \equiv \overline f\gamma_\lambda\gamma^5 f\,.
                \end{eqnarray}

To first order in $G_F$,
                \begin{eqnarray}
\Gamma_\lambda = e \int d^4x\, d^4y\, d^4z\,e^{-ik\cdot x}e^{ik'\cdot y}
\left<  T\,U^{{\rm (em)}} \nu(y) j_\lambda(0) \overline\nu(x)
i {\cal L}_{{\rm int}}^{{\rm (weak)}}(z)\right> _a
\label{1storder}
                \end{eqnarray}
which, using Eq. (\ref{Lweak2}) and amputating the external
neutrino lines, reduces to
                \begin{eqnarray}
\Gamma_\lambda &=& - i eG_F\sqrt 2 {\cal A} \gamma^\rho L \times
\nonumber\\
&& \left( \int{d^4z \, e^{-iq\cdot z} \left<  T \, U^{{\rm (em)}}
j_\lambda(0) j_\rho(z) \right> }
+ \int{d^4z\,e^{-iq\cdot z}\left<  T\,U^{{\rm (em)}}
j_\lambda(0)j_\rho^5(z)\right> } \right)\,.
                \end{eqnarray}
Defining
                \begin{eqnarray}
i\pi^5_{\lambda\rho} & = & (-ie)^2 \int{d^4x e^{iq\cdot x}\left<  T\,
U^{{\rm (em)}}j_\lambda(x)j^5_\rho (0)\right> }\,,\nonumber\\
& = & (-ie)^2 \int{d^4x e^{-iq\cdot x}\left<  T\,
U^{{\rm (em)}}j_\lambda(0)j^5_\rho(x)\right> }\,,
                \end{eqnarray}
we finally obtain the relation
                \begin{eqnarray}
\Gamma_\lambda = - \frac{G_F}{\sqrt 2 e}\gamma^\rho(1 - \gamma^5)
({\cal A}\pi_{\lambda\rho} + {\cal B}\pi^5_{\lambda\rho})\,.
\label{Gampi2}
                \end{eqnarray}
This expression is the same as Eq. (\ref{Gampi}), except that now it
is clear that it is valid in all orders of the electromagnetic interactions.
Since $\pi^5_{\lambda\rho}$ is a pseudotensor that depends only on $q$
and $v$,
it must be proportional to $\epsilon_{\lambda\rho\alpha\beta}q^\alpha
v^\beta$, and therefore it does not contribute to $\Gamma_0$ in the
rest frame of the medium. If one includes the effects of strong
interaction as well, the various occurences of ${\cal L}{\rm
^{(em)}_{int}}$ should be replaced by ${\cal L}{\rm
^{(em)}_{int}} + {\cal L}{\rm ^{(strong)}_{int}}$, but this would not
change Eq. (\ref{Gampi2}).

\section{Expression for the induced charge}
                   The relation between the induced charge and the
Debye screening length is obtained as follows.
As already argued, the zeroth-component of $\Gamma_\lambda$
is given by
                \begin{eqnarray}
\label{gammapirel}
\Gamma_0 (0,\vec q \to 0) = -\frac{G_F{\cal A}}{\sqrt 2 e}\gamma^\rho [1
- \gamma^5] \pi_{0\rho} (0,\vec q \to 0) \,.
                \end{eqnarray}
The most general form of $\pi_{\lambda\rho}$ is \cite{NiPa89b}
                \begin{eqnarray}
\label{pimunugeneral}
\pi_{\lambda\rho} = \pi_T R_{\lambda\rho} + \pi_L Q_{\lambda\rho} +
\pi_P P_{\lambda\rho} \,,
                \end{eqnarray}
where
                \begin{eqnarray}
R_{\lambda\rho} &\equiv & \tilde g_{\lambda\rho} -
Q_{\lambda\rho}\,,\\
Q_{\lambda\rho} &\equiv & \frac{\tilde v_\lambda\tilde v_\rho}{\tilde
v^2}\,, \\
P_{\lambda\rho} &\equiv& {i \over {\cal Q}}
\epsilon_{\lambda\rho\alpha\beta} q^\alpha v^\beta \,,
                \end{eqnarray}
with
                \begin{eqnarray}
\tilde g_{\lambda\rho} &\equiv& g_{\lambda\rho} - \frac{q_\lambda
q_\rho}{q^2} \\
\tilde v_\lambda &\equiv& \tilde g_{\lambda\rho}v^\rho\, \\
{\cal Q} &\equiv & \sqrt{(q\cdot v)^2 - q^2} \,.
                \end{eqnarray}
In the rest frame of the medium, with $v^\mu=(1,0,0,0)$, the above
definitions give
                \begin{eqnarray}
Q_{\lambda\rho}(0,\vec q) & = & v_\lambda v_\rho\,,\\
R_{00}(0,\vec q) = R_{i0}(0,\vec q) = R_{0i}(0,\vec q) & = & 0\,,\\
P_{00}(0,\vec q) = P_{i0}(0,\vec q) = P_{0i}(0,\vec q) & = &
0\,,
                \end{eqnarray}
where we have indicated explicitly the fact that we are
evaluating $R$, $Q$ and $P$ in the static limit, $q^0 = 0$.  From
Eq. (\ref{pimunugeneral}) we then obtain
                \begin{eqnarray}
\pi_{00}(0,\vec q) & = & \pi_L(0,\vec q)\,,\\
\pi_{0i}(0,\vec q) & = & \pi_{i0}(0,\vec q) = 0\,.
                \end{eqnarray}
Eq. (\ref{gammapirel}) then yields
                \begin{eqnarray}
\Gamma_0 (0, \vec q \to 0)= - \left(\frac{G_F{\cal A}}{\sqrt 2
e}\right)\gamma_0 (1 - \gamma_5) \pi_L(0,\vec q\to 0)\,.
                \end{eqnarray}
Using the definition of the Debye screening length,
                \begin{eqnarray}
r_D^{-2} = \pi_L(0,\vec q\to 0)\,,
\label{Debye}
                \end{eqnarray}
we finally obtain for the induced neutrino charge by the use of Eq.
(\ref{def:effch}):
                \begin{eqnarray}
e^{(\nu)}_{{\rm ind}} = - \frac{G_F{\cal A}}{\sqrt 2 er_D^2} (1 -
\lambda) \,,
\label{result}
                \end{eqnarray}
where, as stated before, $\lambda$ is the helicity of the neutrino.
Thus, it is clear that only the left-handed neutrinos have an induced
charge. The  induced charge for the right handed neutrinos vanish
since they have no weak interactions. If they interact via some other
weaker interaction, then of course they also acquire an induced
charge, but the magnitude of that will be further suppressed. Also,
note that $e^{(\nu)}_{{\rm ind}} \propto e$, since
$r_D^2 \propto e^{-2}$ which follows from Eq. (\ref{Debye}).

\section{Generalization to massive neutrinos}
               The generalization of our previous results to massive
neutrinos is straightforward, although there are several important
differences. The effective charge is defined by
Eq. (\ref{matrixeffch}), but the expression for the spinor projection
operator $S(k)$, as well
as the expressison for $\Gamma_0$, differ from the previous case.
The spinor projection operator $S(k)$ is given, for massive particles, by
                \begin{eqnarray}
\label{massiveproj}
S(k) = u(k) \otimes \overline u(k) = \frac{1}{2} (\rlap/ k + m) (1 +
\lambda \gamma^5 \rlap/ s)\,,
                \end{eqnarray}
where $s^\mu$ is the spin polarization vector which, for helicity states,
is given by
                \begin{eqnarray}
s^\mu = {1 \over m} \left( |\vec k| , E \vec k/|\vec k| \right) \,.
                \end{eqnarray}
Therefore, although Eq. (\ref{matrixeffch}) remains valid
for massive neutrinos, Eq. (\ref{def:effch}) does not. To proceed,
we consider the cases of Dirac and Majorana
neutrinos separately.

\subsection{Dirac case}
             For Dirac neutrinos, $\Gamma_\lambda$ remains to be given by
Eq. (\ref{Gampi2}) and the effective charge by Eq. (\ref{matrixeffch}).
The formula for the effective charge of Dirac neutrinos
is now obtained by substituting  Eq. (\ref{massiveproj}) in
Eq. (\ref{matrixeffch}), yielding
                \begin{eqnarray}
e^{(\nu^D)}_{{\rm ind}} = -\frac{G_F{\cal A}}{\sqrt{2}e r_D^2} \left(1 -
\frac{\lambda |\vec k|}{E}  \right) \,.
                \end{eqnarray}

Specializing this formula to the  case of massless neutrinos we recover
Eq. (\ref{result}), as it should be. In the opposite limit of non-relativistic
neutrinos, we see that both helicity states have the same effective charge,
which is equal to half the value of the charge of the left-handed
neutrino in the massless case.

\subsection{Majorana case}
                 For (massive) Majorana neutrinos we again have to use
the spinor projection operator appropriate for massive particles,
given in Eq. (\ref{massiveproj}). However, the formula for
$\Gamma_\lambda$ is modified as follows. Going back to
Eq. (\ref{1storder}), it is important to recognize that each one of
the neutrino field operators $\nu(y)$ and $\overline\nu(y)$ can be
contracted with either one of the same field operators that come from
the factor ${\cal L}{\rm _{int}^{(weak)}}$. The reason is that for a
Majorana neutrino the field operator is self-conjugate and therefore
$\nu(y)$ can be contracted with not only $\overline\nu(y)$ but also
with itself. The net result of adding these two possible contractions
is that the expression for $\Gamma_\lambda$ given in
Eq. (\ref{Gampi2}) is replaced by \cite{MoPaBook}
                \begin{eqnarray}
\label{gammamaj}
\Gamma_\lambda^{(M)}
 = -\frac{G_F}{\sqrt{2}e}(-2\gamma^\rho\gamma^5)
({\cal A}\pi_{\lambda\rho} + {\cal B}\pi^5_{\lambda\rho}) \,.
                \end{eqnarray}
Substituting this formula and the projection operator given above for
the massive case, into Eq. (\ref{matrixeffch}) we then obtain the
effective charge for Majorana neutrinos
                \begin{eqnarray}
e^{(\nu^M)}_{{\rm ind}}
 =  \frac{\sqrt{2} G_F {\cal A}}{er_D^2} \, \frac{\lambda |\vec k|}{E} \,.
                \end{eqnarray}
We notice the following features: ($i$) the positive and negative
helicity states
have opposite effective charge; ($ii$) in the masless limit the result for
the negative helicity state is the same one obtained in the massless
Dirac case and in the Weyl case, while the positive helicity state has the
opposite value of the charge. This is not surprising since the right-handed
component of the Majorana neutrino is just the CPT conjugate of the
left-handed one.

\section{Numerical estimates}
                 In order to obtain numerical estimates, we note that
since $|\vec k|/E \leq 1$, the magnitude of the induced charge is maximum if
the neutrino is massless. Thus, in this section, we use Eq.
(\ref{result}), which is valid for massless neutrinos. We see that we
need to know
two things in order to obtain a numerical estimate for the induced
charge of the neutrino, viz., $\cal A$ and $r_D$. The first is easy,
and is given by the standard model of electroweak interactions. In
fact, one obtains
                \begin{eqnarray}
{\cal A} = \left\{ \begin{array}{ll}
2 \sin^2 \theta_W + {1 \over 2} \qquad & \mbox{for $\nu_e$} \\
2 \sin^2 \theta_W - {1 \over 2}  & \mbox{for $\nu_\mu, \nu_\tau$} \,.
\end{array} \right.
\label{A}
                \end{eqnarray}
For massless $\nu_e$, our result exactly reproduces the results of
Oraevsky and Semikoz \cite{OrSe87}.  Our formulas can be used to
obtain the induced charges for the $\nu_\mu$ and $\nu_\tau$ as well,
even if they are massive particles.

One curious thing to note is the fact that the induced charge of the
$\nu_e$ is different from that of
$\nu_\mu$ or $\nu_\tau$ since the $\nu_e$'s interact with the
electrons of the medium via both charge and neutral currents.
Thus, if neutrinos mix, when they oscillate during their propagation
through a medium, the induced charges also oscillate.
Using $\sin^2 \theta_W = 0.23$, we see that
                \begin{eqnarray}
{e^{(\nu_e)}_{{\rm ind}} \over e^{(\nu_\mu)}_{{\rm ind}}} = -24 \,.
                \end{eqnarray}
It therefore seems that this oscillation of charges should be a fantastic
phenomenon, judging by the fact that the ratio of the induced charges
is large, and also is negative. However, that is not the case because
the magnitude of the induced charge appears to be extremely small. In
fact, putting numbers in Eq. (\ref{result}), one obtains
                \begin{eqnarray}
e^{(\nu_e)}_{{\rm ind}} = -2 \cdot 10^{-32} \times \left( {1\,{\rm
cm} \over r_D} \right)^2 \,.
                \end{eqnarray}
To proceed, we need the value of $r_D$ for the background. This can be
obtained either from the results of Ref. \cite{DNP89} or from standard
texts on plasma physics.
For a background consisting of non-relativistic electrons at
temperature $T$, the Debye radius is given by
                \begin{eqnarray}
r_D^2 = {T \over n_e e^2} \,,
                \end{eqnarray}
where $n_e$ is the electron number density\footnote{Formulas appearing
in books on Plasma Physics usually have a factor $4\pi$ in the
denominator on the right hand side since their definition of electric
charge is different.}. In order for the
induced neutrino charge to be detectable in experiments, the values of
$T$ and $n_e$ must be such that the resulting Debye radius is small
enough. While this is not the case for any known plasma, the methods
presented here may prove useful in applications to similar situations
where more exciting results may be obtained.

\paragraph*{Note added in proof~:} After this work was submitted for
publication, we were made aware of a paper by Altherr and Kainulainen
\cite{AlKa91} where the one-loop electromagnetic vertex of neutrinos has
been calculated in a medium. The calculation agrees with that of Ref.\
\cite{DNP89}. These authors specifically noted that an induced charge
appears in the medium. No effort was made to make contact with the
Debye radius.

\paragraph*{Acknowledgements~:} The work of PBP was supported by a
grant from the Department of Energy.

\newpage

                        \begin{figure}
                \begin{center}
                \begin{picture}(110,125)(-30,35)
\thicklines
\put(85,130){\makebox(0,0){{\huge (a)}}}
\put(-8.5,130){\line(-1,1){15}}
\put(-16.5,138){\vector(-1,1){1}}
\put(-22.5,135){\makebox(0,0){{\large$\nu(k')$}}}
\put(-8.5,130){\line(-1,-1){15}}
\put(-16.5,122){\vector(1,1){1}}
\put(-22.5,125){\makebox(0,0){{\large$\nu(k)$}}}
\multiput(14,130)(-6,0){4}{\oval(3,3)[t]}
\multiput(11,130)(-6,0){4}{\oval(3,3)[b]}
\put(0,136){\makebox(0,0){{\large$Z(q)$}}}
\put(23.5,130){\circle{16}}
\multiput(33,130)(6,0){4}{\oval(3,3)[b]}
\multiput(36,130)(6,0){4}{\oval(3,3)[t]}
\put(57,130){\oval(3,3)[bl]}
\put(58,128.5){\vector(1,0){1}}
\put(50,136){\makebox(0,0){{\large$\gamma(q)$}}}
\put(85,70){\makebox(0,0){{\huge (b)}}}
\put(-8.5,85){\line(-1,1){15}}
\put(-16.5,93){\vector(-1,1){1}}
\put(-22.5,90){\makebox(0,0){{\large$\nu(k')$}}}
\put(-8.5,55){\line(-1,-1){15}}
\put(-16.5,47){\vector(1,1){1}}
\put(-22.5,50){\makebox(0,0){{\large$\nu(k)$}}}
\multiput(-8.5,83.5)(0,-6){5}{\oval(3,3)[l]}
\multiput(-8.5,80.5)(0,-6){4}{\oval(3,3)[r]}
\put(-8.5,56.5){\oval(3,3)[tr]}
\put(-16,70){\makebox(0,0){{\large$W$}}}                
\put(-8.5,85){\line(1,-1){15}}
\put(-0.5,77){\vector(-1,1){1}}
\put(-8.5,55){\line(1,1){15}}
\put(-0.5,63){\vector(1,1){1}}
\multiput(8,70)(6,0){4}{\oval(3,3)[b]}
\multiput(11,70)(6,0){4}{\oval(3,3)[t]}
\put(32,70){\oval(3,3)[bl]}
\put(33,68.5){\vector(1,0){1}}
\put(20,76){\makebox(0,0){{\large$\gamma(q)$}}}
                        \end{picture}
                \end{center}
\caption[]{\sf 1-loop diagrams for the effective electromagnetic
vertex of neutrinos which contribute in the limit that the 4-momenta
of the $W$ and the $Z$ lines are neglected.}\label{f:1loop}
                        \end{figure}

                        \begin{figure}
                \begin{center}
                \begin{picture}(70,25)(-10,20)
\thicklines
\put(17,30){\line(-1,1){15}}
\put(17,30){\line(-1,-1){15}}
\put(9,38){\vector(-1,1){1}}
\put(9,22){\vector(1,1){1}}
\put(3,35){\makebox(0,0){{\large$\nu(k')$}}}
\put(3,25){\makebox(0,0){{\large$\nu(k)$}}}
\put(25,30){\circle{16}}
\multiput(34.5,30)(6,0){4}{\oval(3,3)[b]}
\multiput(37.5,30)(6,0){4}{\oval(3,3)[t]}
\put(58.5,30){\oval(3,3)[bl]}
\put(59.5,28.5){\vector(1,0){1}}
\put(50,36){\makebox(0,0){{\large$\gamma(q)$}}}
                        \end{picture}
                \end{center}
\caption[]{\sf  The diagrams of Fig.~\ref{f:1loop} in the limit of
infinitely heavy $W$ and $Z$ masses.}\label{f:4fermi}
                        \end{figure}
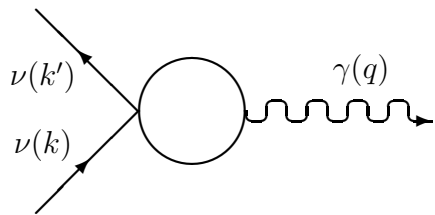

\begin{thebibliography}{[000]}
  \bibitem{palrev} For references and a recent review, see, e.g.,
P B Pal: Int. J. Mod. Phys. A7 (1992) 5387.
  \bibitem{MiSm86} S P Mikheyev, A Y Smirnov: Nuovo Cimento C9 (1986)
17.
  \bibitem{Wol78} L Wolfenstein: Phys. Rev. D17 (1978) 2369.
  \bibitem{NiPa89} J F Nieves, P B Pal: Phys. Rev. D40 (1989) 1693.
  \bibitem{SeSm89} V B Semikoz, Y A Smorodinskii: JETP 68
(1989) 20.
  \bibitem{OrSe87} V N Oraevsky, V B Semikoz: Physica 142A (1987)
135.
  \bibitem{OPSS87} V N Oraevskii, A Y Plakhov, V B Semikoz, Y A
Smorodinskii, JETP 66 (1987) 890.
  \bibitem{DNP89} J C D'Olivo, J F Nieves, P B Pal: Phys. Rev. D40 (1989)
3679.
  \bibitem{DNP90} J C D'Olivo, J F Nieves, P B Pal: Phys. Rev. Lett. 64 (1990)
1088.
  \bibitem{GKL91} C Giunti, C W Kim, W P Lam: Phys. Rev. D43 (1991) 164.
  \bibitem{Saw92} R F Sawyer: Phys. Rev. D46 (1992) 1180.
  \bibitem{GKLL92} C Giunti, C W Kim, U W Lee, W P Lam: Phys. Rev. D45
(1992) 1557.
  \bibitem{ARW63} J B Adams, M A Ruderman, C-H Woo: Phys. Rev. 129 (1963) 1383.
  \bibitem{BrSe93} E Braaten, D Segel: {\sl ``Neutrino energy loss from
plasma process at all temperatures and densities''},  Northwestern
University preprint NUHEP-TH-93-1 (January 1993).
  \bibitem{NiPa89b} J F Nieves, P B Pal: Phys. Rev. D39 (1989) 652, {\bf
40} (1989) 2148(E).
  \bibitem{MoPaBook} For details of this procedure, see e.g., R N
Mohapatra and P B Pal: {\sl Massive neutrinos in Physics and
Astrophysics} (World Scientific, Singapore 1991), Sec. 10.1.2.
  \bibitem{AlKa91} T. Altherr and K. Kainulainen: Phys. Lett. B262
(1991) 79.
                        \end{thebibliography}
\end{document}